\documentclass[11pt]{article}

\usepackage{amsmath, amsfonts}

\newcommand{\Title}[1]{
  ~\vspace{12mm}
  \begin{center}
    \Large\bf #1
  \end{center}
  }
\newcommand{\Author}[1]{
  \begin{center}
    \large #1
  \end{center}
  }
\newcommand{\Institution}[1]{
  ~\vspace{-26pt}
  \begin{center}
   #1
  \end{center}
  }

\newcommand{\History}[1]{
\centerline{\small #1}
}

\newcommand{\ms}{\medskip}
\newcommand{\be}{\begin{equation}}
\newcommand{\ee}{\end{equation}}
\newcommand{\bea}{\begin{eqnarray}}
\newcommand{\eea}{\end{eqnarray}}
\renewcommand{\d}{\ensuremath{\partial}}
\renewcommand{\d}{\ensuremath{\partial}}
\renewcommand{\(}{\ensuremath{\left( }}
\renewcommand{\)}{\ensuremath{\right) }}
\renewcommand{\[}{\ensuremath{\left[}}
\renewcommand{\]}{\ensuremath{\right] }}
\renewcommand{\t}[1]{\widetilde{#1}}
\renewcommand{\a}{\ensuremath{\alpha}}
\renewcommand{\b}{\ensuremath{\beta}}
\newcommand{\g}{{\ensuremath{\gamma}}}

\newcommand{\cO}{\ensuremath{\mathcal{O}}}
\newcommand{\dontprint}[1]{\relax}

\begin{document} \setlength{\unitlength}{1mm}

\thispagestyle{empty}
\rightline{\tt UCB-PTH-09/30}
\Title{Covariant Star Product for Exterior Differential Forms on Symplectic Manifolds\footnote{This work was supported in part by the Director, Office of Science,
Office of High Energy and Nuclear Physics, Division of High Energy
Physics of the U.S. Department of Energy under Contract
DE-AC02-05CH11231, in part by the National Science Foundation under
grant PHY-0457315. }}
\bigskip
\Author{Shannon McCurdy\footnote{smccurdy@berkeley.edu} $\qquad \qquad$ Bruno Zumino\footnote{zumino@thsrv.lbl.gov}}
\Institution{
\it{Center for Theoretical Physics and Department of Physics, University of California\\
Berkeley, CA 94720-7300, USA\\
and\\
Theoretical Physics Group, Lawrence Berkeley National 
Laboratory\\
Berkeley, CA 94720-8162, USA}}
\begin{center}
\ms\ms
\History{October 2$^{nd}$, 2009}
\end{center}
\ms\ms

\begin{abstract}
After a brief description of the $\mathbb{Z}$-graded differential Poisson algebra, we introduce a covariant star product for exterior differential forms and give an explicit expression for it up to second order in the deformation parameter $\hbar$, in the case of symplectic manifolds.  The graded differential Poisson algebra endows the manifold with a connection, not necessarily torsion-free, and places upon the connection various constraints.
 \end{abstract}

\section{Introduction}

A star product is a deformation of an associative, commutative product of functions into a product that is still associative but no longer commutative.  The original motivation behind star products was an alternative theory of quantization; Ref. \cite{f} equated quantization with a deformation of the algebra of classical observables of functions on phase space, where the $\cO(\hbar)$ term in the deformation can be taken to be the classical Poisson bracket (see Ref. \cite{ds} for a historical overview and references therein).  Star products were subsequently found to apply in many areas of physics, including string theory.

From the work of Kontsevich \cite{k}, Cattaneo and Felder \cite{cf1}, and many others, the star product of functions on general Poisson manifolds is well known, in standard coordinates on $\mathbb{R}^d$, to all orders in the deformation parameter. Recently, Ref. \cite{a} wrote the explicit form (to $\cO(\hbar^3)$) of a covariant star product of functions on Poisson manifolds with torsion-free linear connections. 

Motivated by the study of non-commutative gauge theories in string theory, we investigate the deformation of the $\mathbb{Z}$-graded exterior algebra of differential forms\footnote{The $\mathbb{Z}_2$-graded case of superspace is considered in Ref. \cite{cf2}.}.  Since the  $\cO(\hbar)$ term of the star product on functions is the Poisson bracket, we need a generalization of the Poisson bracket to differential forms.  We use the differential Poisson algebra discussed in Ref. \cite{ch} and, more recently, Ref. \cite{Beggs:2003ne}.  Ref. \cite{h} discusses similar material from the perspective of contravariant connections; the approach we use is more general  (see Ref. \cite{Beggs:2003ne}).  As found in Ref. \cite{ch} and Ref. \cite{Beggs:2003ne}, endowing a symplectic manifold with a differential Poisson calculus is equivalent to endowing a symplectic manifold with a connection $\Gamma$ that has certain important restrictions, Equations (\ref{zero_curv})-(\ref{yang_baxter}).  Although these results are already in the literature, we include a brief treatment of the differential Poisson calculus for completeness. 

Modulo the important restrictions due to the differential Poisson calculus, we find that it is possible to define a covariant star product to $\cO(\hbar^2)$ on the space of differential forms on symplectic manifolds endowed with a differential Poisson calculus, and we give the explicit form in this note in Equations (\ref{pbforms}) and (\ref{star_ansatz2}).

\section{The Differential Poisson Calculus \label{diffpcalc} }
For all exterior differential forms\footnote{A differential form of degree $|\a|$ is a covariant tensor field of degree $|\a|$ that is totally antisymmetric.  In a local coordinate system $\{ x^i, \dots, x^n\}$,  a $|\a|$-form $\a$ can be uniquely expressed as $ \a =\tfrac{1}{|\a|!} \a_{i_1 \dots i_{|\a|}} dx^{i_1} \wedge \dots \wedge dx^{i_{|\a|}}$ where the coefficient $ \a_{i_1 \dots i_{|\a|}} $ is completely antisymmetric.} $\a, \b,$ and $\g$, a graded differential Poisson algebra on a Poisson manifold must satisfy the following axioms:
\bea
& |\{\a,\b\}|=|\a|+|\b|&  \label{BD} \\
&\{\a,\b\}=(-1)^{|\a||\b|+1}\{\b,\a\} & \\
& \{\a,\b+ \g\}=\{\a,\b\}+\{\a,\g\} \label{additivity} & \\
& \{\a,\b\wedge\g\}=\{\a,\b\}\wedge\g+(-1)^{|\a||\b|}\b\wedge\{\a,\g\}  \label{gradedproduct} & \\
&\{\a,\{\b,\g\}\} +(-1)^{|\a|(|\b|+|\g|)}\{\b,\{\g,\a\}\}+(-1)^{|\g|(|\a|+|\b|)}\{\g,\{\a,\b\}\}=0&  \label{Graded}  \\
& d\{\a,\b\}=\{d\a,\b\}+(-1)^{|\a|}\{\a,d\b\}. \label{leibnizrule} &
\eea
Note that $d^2=0$, and we take $d$ to be undeformed.  Axioms (\ref{BD})-(\ref{Graded}) give a $\mathbb{Z}$-graded Poisson bracket.  The graded Jacobi identity (\ref{Graded}) and the Leibniz rule (\ref{leibnizrule}) constrain the Poisson bracket on differential forms.  In particular, on a symplectic manifold, the differential Poisson bracket between a function $f$ and a $1$-form $\a$ has all of the properties of a covariant differential operator in each argument, and thus defines a connection on the symplectic manifold,
\be
\nabla_{X_f} \a  \equiv \{ f, \a\} , \nonumber
\ee
where $X_f$ is the Hamiltonian vector field associated with $f$.  The Leibniz rule (\ref{leibnizrule}) relates the coordinate variation of the Poisson bivector to the connection coefficients.  Consider 
\be
d\{ f, g \} = \{d f, g\} +  \{ f, dg\}, \label{leibniz1}
\ee
where $f,g$ are both functions.  Since this must be true for all functions $f,g$, the Leibniz rule (\ref{leibniz1}) becomes, in local coordinates,
\be
\d_k \pi^{ij} =- \pi^{mj}\Gamma_{mk}^i-\pi^{im}\Gamma_{mk}^j,  \label{leibnizcoords}
\ee
where $\pi^{ij}$ is the Poisson bivector (with inverse $\omega_{ij}$ on symplectic manifolds), and $\Gamma_{mk}^i$ are the connection coefficients.  In terms of the covariant derivative and the torsion, (\ref{leibnizcoords}) can also be rewritten as $\nabla_k \pi^{ij} = -\pi^{ir} T^j_{rk} - \pi^{rj}T^i_{rk}. $  From the same connection coefficients $\Gamma_{mk}^i$, one can define two connections $\nabla, \t{\nabla}$ with respective curvatures $R, \t{R}$.  For example,
\be
\nabla_i dx^k  =-\Gamma_{ij}^kdx^j  \qquad \qquad \t{\nabla}_idx^k =  -\Gamma_{ji}^kdx^j.\label{twoconnections}
\ee
These connections and curvatures differ when the torsion is non-zero.  In this convenient notation, discussed further in the Appendix, the Leibniz rule (\ref{leibnizcoords})  becomes
\be
\t{\nabla}_k \pi^{ij} =0. \label{sc}
\ee
Thus $\pi^{ij}$ is covariantly constant under $\t{\nabla}$, and $\t{\nabla}$ is an
almost symplectic connection\footnote{In the literature, symplectic
connections are additionally taken to be torsionless, but the connection
defined here may have torsion.}.  

One can use the Leibniz condition (\ref{leibnizcoords}) together with the Jacobi identity for the Poisson bivector to arrive at a cyclic relation for the torsion:
\be
\sum_{(i,j,k)} \pi^{i\ell}\pi^{jm}T_{\ell m}^k=0. \label{cycT}
\ee
Note that a torsion-free connection is sufficient but not necessary to satisfy (\ref{cycT}). 
Also note that (\ref{leibnizcoords}) and the Jacobi identity for the Poisson bivector can also be combined to obtain the following cyclic relation:
\be
\sum_{(i,j,k)} \pi^{i\ell}\nabla_\ell \pi^{jk}=0 \label{cycnabla}.
\ee
As before, note that a symplectic connection is sufficient but not necessary to satisfy (\ref{cycnabla}).  If, in addition to  $\t{\nabla}_k \pi^{ij}=0$, one imposes $\nabla_k \pi^{ij}=0$, then $T_{ij}^k =0$ and there is only one covariant derivative $\nabla=\t{ \nabla} $.  Ref. \cite{Beggs:2003ne} has a nice proof.

In this note, we do not require $\nabla_k \pi^{ij}=0$.

To find the local coordinate expression for Poisson bracket between two $1$-forms, consider the Leibniz rule again:
\bea
\{ df, dg \}=d \{f , dg\}  =d \( \pi^{ij}\d_i f \nabla_j dg \) =  \pi^{ij} \nabla_i df \wedge \nabla_j dg  - \t{R}^{ij} \wedge i_i df \wedge i_j dg, \label{leibniz2}
\eea
where, to get the right-most equality of (\ref{leibniz2}), we have introduced a contraction operator\footnote{The contraction operator $i_k$ is discussed fully in the Appendix. The interior product of a form $\a$ with a vector $X$ has the following relation to the contraction operator: $i_X \a = X^k i_k \a$.} $i_k$ and defined $\t{R}^{ij} \equiv \tfrac12 \pi^{ik} \t{R}^{j}_{kab} dx^a \wedge dx^b$, where $\t{R}^{j}_{kab}$ are the components of the curvature of $\t{\nabla}$.  An important property $\t{R}^{ij}$  is its symmetry in the upper two indices,  $\t{R}^{ij}=  \t{R}^{ji}$.  This is easily shown from  $[\t{\nabla}_m, \t{\nabla}_n]\pi^{ij}=0$.

Using the graded product rule, one arrives at the general form of the Poisson bracket between differential forms:
\be
\{ \a, \b \}= \pi^{ij} \nabla_i \a \wedge \nabla_j \b  +(-1)^{|\a|} \t{R}^{ij} \wedge i_i \a \wedge i_j \b. \label{pbforms}
\ee
However, (\ref{pbforms}) does not satisfy the graded Jacobi identity a priori.  Using (\ref{pbforms}) and the identities $\t{R}^{ij}=  \t{R}^{ji},$ 
(\ref{ii}), and (\ref{inabla}), as well as the following two identities\footnote{The brackets $\{, \}$ in (\ref{nablanabla}) are anticommutators.}
\bea
 \nabla_i \nabla_m \a &=& \tfrac12 \{ \nabla_i, \nabla_m\} \a +\tfrac12 \[ \nabla_i, \nabla_m\] \a \label{nablanabla}\\
\[ \nabla_i ,\nabla_m \] \a &=& - R^p_{aim}dx^a \wedge i_p \a -T^p_{im} \nabla_p \a, \label{commutator}
\eea
 we find: 
\begin{eqnarray}
\lefteqn{ \{\a, \{\b, \g\} \}+(-1)^{|\a|(|\b| + |\g|)}\{\b, \{\g, \a\}\}- \{ \{\a, \b\}, \g\}= }  \label{explicitjacobi}\\
 && \big( \pi^{mn} \d_n \pi^{ij}+  \pi^{in} \d_n \pi^{jm} +  \pi^{jn} \d_n \pi^{mi} \big) \nabla_i \a \wedge \nabla_j \b \wedge \nabla_m \g  \nonumber \\
&&- \pi^{ij} \pi^{mn} R^p_{aim} dx^a \wedge \big( i_p \a \wedge \nabla_j \b \wedge \nabla_n \g + \nabla_n \a \wedge i_p \b \wedge \nabla_j\g + \nabla_j \a \wedge \nabla_n \b \wedge i_p \g  \big) \nonumber \\
   &&+ \pi^{mn}  \nabla_n \t{R}^{ij} \wedge \big( (-1)^{|\a|}  i_i \a \wedge i_j \b \wedge \nabla_m \g +(-1)^{|\b|} \nabla_m \a \wedge i_i \b \wedge i_j \g \nonumber\\
   &&- (-1)^{|\a| + |\b|}  i_j \a \wedge\nabla_m \b\wedge i_i \g   \big)\nonumber\\
 && -(-1)^{|\b|} \big( \t{R}^{ij} \wedge i_i \t{R}^{mn}+\t{R}^{im} \wedge i_i \t{R}^{nj}+\t{R}^{in} \wedge i_i \t{R}^{jm}\big) \wedge  i_m \a \wedge i_n \b \wedge i_j \g.  \nonumber
\end{eqnarray}
For this to vanish for all $\a, \b, \g$, the connection coefficients $\Gamma^i_{jk}$ must satisfy several additional conditions\footnote{The
classical Yang-Baxter equation (CYBE) is given by $\[r_{12},r_{13}\] +
\[r_{12},r_{23}\]+ \[r_{13},r_{23}\]= 0$ in tensor product notation, where $\[, \]$ is the matrix
commutator Ref. \cite{jf}.  Using  $\t{R}^{ij}_{ab}= \t{R}^{ji}_{ab} =- \t{R}^{ij}_{ba}$, one can verify that (\ref{yang_baxter}) implies that
$\t{R}^{ij}_{ab}$ satisfies the CYBE.}: 
\bea 
&R^i_{jk\ell}=0& \label{zero_curv} \\
&\nabla_a\t{R}^{mn}_{cd}=0&\label{cov_const} \\
 &\t{R}^{ab}\wedge i_b \t{R}^{mn}+\t{R}^{mb}\wedge i_b\t{R}^{na}+\t{R}^{nb}\wedge i_b \t{R}^{am}=0. \label{yang_baxter}
\eea 
Due to the Leibniz rule, these conditions are not independent.  For example, consider three functions, $f, g,$ and $h$.  Following the argument in Ref. \cite{Beggs:2003ne}, let us define four functions:
\bea
J_0(f,g,h)&=&  \{f,\{g,h\}\} + \{g,\{h,f\}\}+\{h,\{f,g\}\} \label{J_0} \nonumber \\
J_1(f,g,h)&=&   \{f,\{g,dh\}\} + \{g,\{dh,f\}\}+\{dh,\{f,g\}\}   \label{J_1} \nonumber \\
J_2(f,g,h)&=&  \{f,\{dg,dh\}\} + \{dg,\{dh,f\}\}-\{dh,\{f,dg\}\}  \label{J_2}  \nonumber\\
J_3(f,g,h)&=&  \{df,\{dg,dh\}\} + \{dg,\{dh,df\}\}+\{dh,\{df,dg\}\}.  \label{J_3}  \nonumber
\eea
These functions $J_i(f, g,h)$ are obstructions to the graded Jacobi identity.  $J_0(f,g,h)=0$ gives the Jacobi identity for the Poisson bivector.  Note that because the Leibniz rule holds, 
\be 
d J_0(f,g,h) = J_1(f,g,h)+ J_1(g,h,f)+J_1(h,f,g).\nonumber
\ee
So, $J_0(f,g,h)=0$ implies that the cyclic permutation of $J_1(f,g,h)$ is identically zero.  Similarly, 
\be
d J_1(f,g,h) =J_2(f,g,h) -J_2(g, h ,f) ,\nonumber
\ee
and, due to the definition, $J_2(f,g,h) = J_2(f,h, g)$, we have, 
\be
d J_1(f,g,h) =J_2(f,g,h) -J_2(g, f , h ).\nonumber
\ee
When $J_1(f,g,h)=0$, this implies that $J_2(f,g,h)$ must be completely symmetric in its arguments.  Finally, 
\be
 d J_2(f,g,h) = J_3(f,g,h),\nonumber
 \ee
so if  $J_2(f,g,h)=0,$ then $J_3(f,g,h)=0$.  Therefore, (\ref{yang_baxter}) contains no new constraints; (\ref{yang_baxter})  is implied by the Leibniz rule, the Jacobi identity for the Poisson bivector, the vanishing of $R$, and the covariant constancy of $\t{R}^{ij}$.

If, in addition to obeying the Leibniz condition and (\ref{zero_curv})-(\ref{yang_baxter}), the connection happens to be torsionless, then $\t{R}$ is identically zero.  This is obvious from Equation (\ref{rzerotr}) in the Appendix.  In the torsionless case, the Poisson bracket reduces to
\be
\{\a,\b\} |_{T=0} = \pi^{mn}\nabla_m\a \wedge \nabla_n\b. \nonumber
\ee

\section{A Covariant Star Product for Exterior Differential Forms}

On symplectic manifolds $M$, we deform $\Omega^*(M)$ to $\Omega^*(M)[[\hbar]]$.  Elements of $\Omega^*(M)[[\hbar]]$ are formal power series in $\hbar$ with coefficients in $\Omega^*(M)$: $\a = \a_0 + \sum_{n=1}^\infty \hbar^n \a_n$.  A product of differential forms in $\Omega^*(M)[[\hbar]]$ is a ``covariant star product of differential forms'' if it satisfies the following properties for $ \a, \b, \g \in\Omega^*(M)[[\hbar]]$:
\begin{enumerate}
\item The product takes the form: $\a*\b=\a\wedge\b + \sum_{n=1}^\infty\hbar^n C_n(\a, \b)$
where the $C_n$ are covariant bilinear differential operators of at most order $n$ in each argument.  The $C_n$ are polynomials of order $n$ in the Poisson bivector.  \label{sum}
\item The product is associative: $\(\a*\b\)*\g = \a*\(\b*\g\)$ \label{star_assoc}
\item The order $\hbar$ term is the differential Poisson bracket for forms (\ref{pbforms}). \label{star_pb}
\item The constant function, $1$, is the identity: $1*\a = \a*1 = \a$
\item The $C_n$ have degree zero:  $|C_n(\a,\b)| = |\a| + |\b|$. \label{stardegree} 
\end{enumerate}  
Condition \ref{star_pb} means that the symplectic manifold $M$ is endowed with a connection, $\Gamma$, that obeys the requisite restrictions (\ref{sc}) and (\ref{zero_curv})-(\ref{yang_baxter}).  Note that we require the Leibniz rule only at $\cO(\hbar)$ of the star product.  

\subsection{Hochschild Cohomology and Associativity}
The associativity condition at $\mathcal{O}(\hbar^n)$ can be expressed as a condition on the Hochschild coboundary of $C_n$; this is well-known in the literature, but we review it briefly for completeness.
For a $\mathbb{Z}$-graded associative algebra $A=\oplus_{j\in \mathbb{Z}} A^j$, where $A^j$ is homogeneous of degree $j$, let $\a_i \in A^{|\a_i|}$.  If $C$ is a Hochschild $p$-cochain, it is $p$-linear in $\a_1,\dots,\a_p$ and homogeneous of degree $|C|$ such that  $C(\a_1, \dots, \a_p) \in A^{|C| + |\a_1| + \dots + |\a_p| }$.  The Hochschild coboundary of $C$, $\delta_H C$, is a $(p+1)$-cochain.  For a $p$-cochain $C$ of degree $|C|$,  $\delta_H C$ is given by:
 \bea
 (\delta_H C)(\a_0, ..., \a_p) &\equiv& (-1)^{|C| |\a_{0}|}\a_0 C(\a_1,...,\a_p) +\sum^{p-1}_{j=0} (-1)^{j+1} C(\a_0, \dots, \a_j \a_{j+1}, \dots ,\a_p) \nonumber \\
 &&+ (-1)^{p+1} C(\a_0, ...,\a_{(p-1)})  \a_p \nonumber
 \eea
A Hochschild $p$-cochain $C$ is called a Hochschild $p$-cocycle if $(\delta_H C)=0$.   Like all coboundary operators, $\delta_H^2=0$.  For $Z_H^p(A, A)$ the space of $p$-cocycles, and $B_H^p(A, A)$ the space of $p$-cocycles that are coboundaries of $(p-1)$-cochains, the Hochschild cohomology space is $H_H^p(A, A)= Z_H^p(A, A)/B_H^p(A, A)$.

The $C_n$ in the star product are Hochschild $2$-cochains of degree zero.  The coboundary of a Hochschild $2$-cochain of degree zero is:
 \be
  (\delta_H C_n)(\a,\b,\g)=\a \wedge C_n (\b, \g) - C_n( \a\wedge\b, \g) + C_n(\a, \b\wedge\g) - C_n(\a, \b) \wedge\g. \nonumber
  \ee
The associativity condition at $\mathcal{O}(\hbar^n)$ can be written as:
\be
 (\delta_H C_n)(\a,\b,\g)= \sum_{r+s=n; r,s > 0} \( C_r(C_s(\a,\b), \g) - C_r(\a, C_s(\b, \g))\) \qquad \forall n \geq 1. \label{hochassoc}
 \ee
Note that the right-hand side of (\ref{hochassoc}) is a 3-cocycle, since $\delta_H^2 C_n= 0$.  Obstructions to extending the deformation are in $H_H^3( \Omega^*(M), \Omega^*(M)),$ the space of Hochschild $3$-cocycles that are not coboundaries.

\subsection{Chevalley Cohomology and Associativity}
Chevalley cohomology has implications for both associativity and equivalence; this is well-known in the literature, but we review it briefly for completeness.  For a $\mathbb{Z}$-graded associative algebra $A=\oplus_{j\in \mathbb{Z}} A^j$, where $A^j$ is homogeneous of degree $j$, let $\a_i \in A^{|\a_i|}.$  Let $\sigma$ denote a permutation of $\{1, \dots, p\}$, let $\epsilon(\sigma)$ denote the sign of the permutation, and let $\epsilon_{|\a|}( \sigma)$ denote the sign of $\sigma$ acting on $\{\a_1, \dots, \a_p\}$ in the graded sense.  If $C$ is $p$-linear in $\a_1,\dots,\a_p$, one says that $C$ is symmetric (respectively antisymmetric) in $\a_1,\dots,\a_p$ if,
\bea
C(\a_{\sigma(1)},\dots,\a_{\sigma(p)})&=&\varepsilon_{|\a|}(\sigma)C(\a_1,\dots,
\a_p)\cr
\Big(\hbox{respectively}~~C(\a_{\sigma(1)},\dots,\a_{\sigma(p)})&=& \varepsilon
(\sigma)\varepsilon_{|\a|}(\sigma)C(\a_1,\dots,\a_p)\Big). \notag
\eea
This is equivalent to, for all $i$,
\bea
C(\a_1,\dots,\a_{i+1},\a_i,\dots,\a_p)&=&(-1)^{|\a_i||\a_{i+1}|}C(\a_1,\dots,\a_p)
\cr
\big(\hbox{respectively}~~C(\a_1,\dots,\a_{i+1},\a_i,\dots,\a_p)&=&-(-1)^{|\a_i|
|\a_{i+1}|}C(\a_1,\dots,\a_p)\big). \notag
\eea

Let the $\mathbb{Z}$-graded associative algebra $A$ be equipped with a  $\mathbb{Z}$-graded Poisson bracket $\{,\}$.  If $C$ is a Chevalley $p$-cochain, it is $p$-linear, antisymmetric, and homogeneous of degree $|C|$ such that $C(\a_1, \dots, \a_p) \in A^{|C| + |\a_1| + \dots + |\a_p| }$.  The Chevalley coboundary $\delta_C C$ of $C$ is $(p+1)$-linear and antisymmetric:
\begin{align*}
(\delta_C&C)(\alpha_0,\dots,\alpha_p)=\cr&\sum_{i=0}^p(-1)^i\varepsilon_{|\alpha|}(i,0\dots\hat{i}\dots p)(-1)^{|C| |\a_i|}\{\alpha_i,
C(\alpha_0,\dots\widehat{\alpha_i}\dots,\alpha_p)\} \cr
&+\sum_{0\leq i < j \leq p}\varepsilon_{|\a|}(i,j,0,\dots\hat{i}
\dots\hat{j}\dots p)(-1)^{i+j }C\left( \{\alpha_i,\alpha_j \} ,\alpha_0,\dots
\widehat{\alpha_i}\dots\widehat{\alpha_j}\dots,\alpha_p\right),
\end{align*}
and $\delta_C^2 = 0$. A Chevalley $p$-cochain is called a Chevalley $p$-cocycle if $(\delta_C C)=0$.  For $Z_C^p(A, A)$ the space of $p$-cocycles, and $B_C^p(A, A)$ the space of $p$-cocycles that are coboundaries of $(p-1)$-cochains, the Chevalley cohomology space is $H_C^p(A, A)= Z_C^p(A, A)/B_C^p(A, A)$.

The antisymmetric part of the associativity relation (\ref{hochassoc}) at $\mathcal{O} (\hbar^3)$ requires that the antisymmetric part of $C_2(\a, \b)$, $C^{-}_2(\a, \b)$, be a Chevalley cocycle:
\be
0= Skew((\delta_H C_3)(\a, \b, \g)) =- 4 (\delta_C C^{-}_2) (\a,\b, \g). \label{chevreq}
\ee
Note that this argument could be extended to $\mathcal{O} (\hbar^k)$.  

\subsection{Star Product Equivalence}

Two star products, $*$ and $\overline{*}$ with  $C_i(\a, \b)$ and $\overline{C}_i(\a, \b)$ respectively, are formally equivalent if there exists a differential operator $T$ of degree $|T|=0$ of the form $T= id + \sum_{n=1}^\infty\hbar^n T_n$ such that
\be
T\(\a\overline{*} \b \)= T\(\a\) * T\(\b\).\nonumber
\ee
The equivalence condition at $\mathcal{O} (\hbar^n)$ is:
\bea
 (\delta_H T_n)(\a, \b) &=& \sum_{r+s=n; s<n} T_s(\overline{C}_r(\a,\b))-  \sum_{r+s+t =n; s,t< n} C_r (T_s(\a), T_t(\b)). \label{equivcond}
\eea
If two star products are equivalent to $\mathcal{O} (\hbar^n)$, then the condition for extending the equivalence is that a Hochschild 2-cocyle constructed from the $T_k$,  $k \leq n $ must be a Hochschild coboundary of $T_{n+1}$.  Note that the right-hand side of (\ref{equivcond}) is a Hochschild $2$-cocycle.  Obstructions to equivalence are in $H_H^2( \Omega^*(M), \Omega^*(M))$.

If two star products, $C_i(\a, \b)$ and $\overline{C}_i(\a, \b)$, are the same at $\mathcal{O} (\hbar)$ and equivalent to $\mathcal{O} (\hbar^2)$, then the associativity relation (\ref{hochassoc}) at $\mathcal{O} (\hbar^2)$ requires that 
\be
(\delta_H \overline{C}_2)(\a, \b, \g)=(\delta_H C_2)(\a, \b, \g). 
\ee
The difference between  $\overline{C}_2(\a, \b)$ and $C_2(\a, \b)$ must be a Hochschild cocycle, which can be parameterized the following way: 
\be
\overline{C}_2(\a, \b) = C_2(\a, \b) +  C^{-}_2(\a, \b) + C^{+}_2(\a, \b) + (\delta_H T_2)(\a, \b), \notag
\ee
where $ C^{-}_2(\a, \b) $ is an antisymmetric Hochschild $2$-cocycle that is an derivation in each argument and $C^{+}_2(\a, \b)$ is a symmetric Hochschild $2$-cocycle that is not a Hochschild coboundary.  The symmetric Hochschild coboundary part is given by $(\delta_H T_2)(\a, \b)$.  Also, $ C^{-}_2(\a, \b)$ must also be a Chevalley cocycle, due to (\ref{chevreq}). 

If $\overline{C}_2(\a, \b)$ and $C_2(\a, \b)$ are to be equivalent star products at $\mathcal{O} (\hbar^2)$ , then (\ref{equivcond}) requires that, for
\be
T(\a) = \a + \hbar T_1 (\a) + \hbar^2\( -  T_2( \a) + \frac12 T_1(T_1(\a)) \)+ \dots, \notag
\ee
$T_1(\a)$ must be a Hochschild cocycle, $C^{+}_2(\a, \b)=0$ , and $ C^{-}_2(\a, \b)$ must be a Chevalley coboundary of $T_1$:
\be
 C^{-}_2(\a, \b) =  - (\delta_C T_1) (\a, \b).  \notag
\ee 
Note that this argument can be extended to $\mathcal{O} (\hbar^k)$.

Thus, antisymmetric obstructions to equivalence at $\mathcal{O} (\hbar^2)$ are given by $H_C^2( \Omega^*(M), \Omega^*(M))$, and symmetric obstructions are given by symmetric terms in $H_H^2( \Omega^*(M), \Omega^*(M))$.

\subsection{Explicit Form of $\a*\b$ to $\mathcal{O}\(\hbar^2\)$}

A star product which satisifies Properties \ref{sum}-\ref{stardegree} to $\mathcal{O}\(\hbar^2\)$ for two arbitrary forms $\a, \b$ has the following $C_2(\a,\b))$:
\bea
\lefteqn{ C_2(\a,\b) \equiv \tfrac12 \pi^{ij}\pi^{mn}\nabla_i\nabla_m\a\wedge\nabla_j\nabla_n\b + \tfrac13( \pi^{mn} \nabla_n \pi^{ij}}  \label{star_ansatz2}\\
&& + \tfrac12 \pi^{in} \pi^{jp} T^m_{np})\big( \nabla_m \nabla_i \a \wedge \nabla_j \b - \nabla_i \a \wedge \nabla_m \nabla_j \b \big) + (-1)^{|\a|} \pi^{ij}\t{R}^{mn}\wedge \nabla_i  i_m \a \wedge \nabla_j  i_n \b \nonumber \\
&&-\tfrac12 \t{R}^{ij}\wedge\t{R}^{mn}\wedge  i_i  i_m \a\wedge i_j  i_n \b  -\tfrac13 \t{R}^{i\ell}\wedge  i_\ell \t{R}^{mn} \wedge\((-1)^{|\a|} i_i i_m \a \wedge  i_n \b +  i_m \a  \wedge i_i i_n \b \). \nonumber
\eea
Properties \ref{sum} and \ref{star_pb}-\ref{stardegree} are manifestly
satisfied by (\ref{pbforms}) and (\ref{star_ansatz2}); a lengthy calculation verifies (\ref{pbforms}) and (\ref{star_ansatz2}) also satisfy Property
\ref{star_assoc}, given by Equation (\ref{hochassoc}):
\bea
\lefteqn{(\delta C_2) (\a, \b, \g)= C_1(C_1(\a,\b),\g) - C_1(\a,C_1(\b,\g)) } \label{assoc2} \\
 &=&+ \tfrac12\pi^{ij} \pi^{mn} \big( \{\nabla_i, \nabla_m \} \a \wedge \nabla_n \b \wedge \nabla_j \g - \nabla_m \a \wedge \nabla_i \b \wedge \{\nabla_n, \nabla_j \} \g \big) \nonumber \\
 && -( \pi^{mn} \nabla_n \pi^{ij} +  \tfrac12\pi^{in} \pi^{jp} T^m_{np})\nabla_i \a \wedge \nabla_m \b \wedge \nabla_j \g  \nonumber\\
 && + \pi^{mn} \t{R}^{ij}\big( (-1)^{|\a|}\nabla_m i_i \a \wedge i_j 
 \b\wedge \nabla_n \g + (-1)^{|\a| + |\b|}i_i \nabla_m 
 \a \wedge \nabla_n \b \wedge i_j \g  \nonumber \\
&& -(-1)^{|\b|} \nabla_m \a \wedge  i_i \b \wedge\nabla_n i_j \g  -(-1)^{|\a| + |\b|  }i_i \a \wedge  \nabla_m \b \wedge  i_j\nabla_n\g\big) \nonumber \\
 && -(-1)^{|\b|} \t{R}^{ij}\wedge i_i \t{R}^{mn}\wedge  i_n \a \wedge i_j \b \wedge i_m \g  \nonumber \\
  && + \t{R}^{ij} \wedge \t{R}^{mn}\wedge \big( (-1)^{|\b|} i_i i_m \a \wedge i_n \b\wedge i_j \g -(-1)^{|\a|+ 1}  i_m \a  \wedge i_i \b \wedge i_n i_j \g \big), \nonumber
\eea
where the brackets $\{, \}$ in (\ref{assoc2}) are anticommutators.  To arrive at this result, we have used Equations (\ref{cycT}), (\ref{cycnabla}), $\t{R}^{ij} = \t{R}^{ji}$, (\ref{zero_curv}), (\ref{cov_const}) and (\ref{yang_baxter}), and the identities (\ref{ii}), (\ref{inabla}), (\ref{nablanabla}), and (\ref{commutator}).   Indeed, (\ref{star_ansatz2}) was constructed by ansatz to satisfy (\ref{assoc2}).

Note that if the connection on $M$ is torsionless, then the connection $\nabla$ is a flat symplectic connection, and the star product reduces to 
\be
\a*\b |_{T=0} = \a \b +\hbar \pi^{mn}\nabla_m\a\wedge\nabla_n\b +\hbar^2  \tfrac12 \pi^{ij}\pi^{mn}\nabla_i\nabla_m\a\wedge\nabla_j\nabla_n\b + \mathcal{O}\(\hbar^3\). \nonumber
\ee
In flat space with $\a, \b$ restricted to functions, this reduces to the first three terms of the case considered in Theorem $5$ of Ref. \cite{f} {\it(no. $1$)}.  Indeed, if the connection is a flat symplectic connection, then the following product between differential forms will be associative to all orders:
\be
\a*\b |_{T=0} = \sum^{\infty}_{r = 0} \tfrac{\hbar^r}{r!} \pi^{i_1 j_1} \dots\pi^{i_r j_r} \nabla_{i_1} \dots\nabla_{i_r} \a \wedge  \nabla_{j_1} \dots\nabla_{j_r} \b. \nonumber
\ee

\section{Conclusion}

In this note, we present a non-commutative deformation (the star product) of
the graded algebra of exterior differential forms to $\mathcal{O}\(\hbar^2\)$.   We review and then use the graded differential Poisson bracket  (\ref{pbforms}) at  $\mathcal{O}\(\hbar\)$, and at $\mathcal{O}\(\hbar^2\)$ we give (\ref{star_ansatz2}).  We verify that this star product satisifes all the necessary
properties, such as associativity, to $\mathcal{O}\(\hbar^2\)$.

The results of this paper may be generalized in at least four directions.  One could examine the Chevalley and Hochschild cohomologies and determine if this star product is unique up to equivalence at $\mathcal{O}\(\hbar^2\)$ and if there are obstructions to extending the star product to $\mathcal{O}\(\hbar^3\)$.  Barring possible obstructions, it should be possible (albeit laborous) to find the explicit star product for differential forms to $\cO(\hbar^3)$ using the associativity relation (\ref{hochassoc}).  These calculations are in progress.  Thirdly, one could look for a proof that this formal star product exists to all orders.  Finally, one could apply this star product between differential forms to physics, such as gauge theories on noncommutative spaces, or generalizing the Seiberg-Witten map \cite{sw}.

\section{Acknowledgments}
The authors thank Anthony Tagliaferro, Dmitri Vassilevich, and Alan Weinstein for useful discussions.  This work was supported in part by the Director, Office of Science, Office of High Energy and Nuclear Physics, Division of High Energy Physics of the U.S. Department of Energy under Contract No. DE-AC02-05CH11231, in part by the National Science Foundation under grant PHY-0457315.

\section{Appendix}
\subsection{The Contraction Operator}
The operation $ i_{j_m}\a$, for $\a$ an $|\a|$-form, denotes:
\bea
i_{j_m}\a&\equiv& \tfrac{1}{|\a|!} \sum^{|\a|}_{m=1} (-1)^{m+1}  \a_{j_1 \dots j_m \dots j_{|\a| }} dx^{j_1}\wedge \dots\wedge \widehat{dx^{j_m}}\wedge \dots \wedge dx^{j_{|\a| }}, \label{contop} 
\eea
where $\widehat{dx^{j_m}}$ means to omit $dx^{j_m}$.  Note that the interior product of  a vector $X $ and $\a$ an $|\a|$-form can be denoted in coordinates using this contraction operator: $i_X \a = X^m i_m \a$.
Similarly, $ i_{n} i_{m}\a  $, for $\a$ an $|\a|$-form, denotes:
\be
i_{n} i_{m}\a = \tfrac{1}{ (|\a|-2)!} \a_{m n j_3 \dots j_{|\a| }} dx^{j_3} \wedge \dots \wedge dx^{j_{|\a| }}= -i_{m} i_{n}\a  \label{ii}.
\ee
And, applying (\ref{contop}) to a wedge product, we see that the contraction operator is an anti-derivation:
\be
i_m (\a\wedge \b) \equiv i_m \a_ \wedge \b + (-1)^{|\a|} \a \wedge i_m \b.
\ee
Furthermore,
\be
\nabla_m i_n \a \equiv \tfrac{1}{ (|\a|-1)!}  \nabla_m \a_{n j_2 \dots j_{|\a| }} dx^{j_2} \wedge \dots \wedge dx^{j_{|\a| }} = i_n \nabla_m  \a . \label{inabla}
\ee 

 Note also that we take the wedge product to act on objects that are not strictly differential forms, like $\nabla_m \a$ and $  i_m \a $.  Objects such as $\nabla_m \a \wedge \nabla_n \b$  
 mean:
 \begin{eqnarray*}
\nabla_m \a \wedge \nabla_n \b &=& \tfrac{1}{|\a|!|\b|!}(\nabla_m \a)_{i_1 ... i_{|\a|}}(\nabla_n \b)_{j_1 ... j_{|\b|}} dx^{i_1} \wedge ... \wedge dx^{i_{|\a|}}\wedge dx^{j_1} \wedge ... \wedge dx^{j_{|\b|}} 
\end{eqnarray*}
where the $m,n$ indices are not antisymmetrized with the $i, j$ indices.

\subsection{The connections $\nabla,\t{\nabla}$ \label{2connections}}

We define two connections from the same connection coefficients $\Gamma_{ij}^k$, as in (\ref{twoconnections}).
The curvature for these two connections are:
\bea
R^i_{mab} &= &
\d_a\Gamma^i_{bm}-\d_b\Gamma^i_{am}+\Gamma^i_{a \ell}\Gamma^\ell_{bm}-\Gamma^i_{b \ell}\Gamma^\ell_{am} \nonumber \\
\t{R}^i_{mab}&= &
\d_a\Gamma^i_{mb}-\d_b\Gamma^i_{ma}+\Gamma^i_{\ell a}\Gamma^\ell_{mb}-\Gamma^i_{\ell b}\Gamma^\ell_{ma}. \nonumber
\eea
The torsion $T^a_{ij} = \Gamma^a_{ij} - \Gamma^a_{ji}$.

As noted in Ref. \cite{ch}, the difference between these two curvatures can be written as:
\be
\t{R}^a_{kij} - R^a_{kij} = - \nabla_i T^a_{jk} - \nabla_j T^a_{ki} +T^a_{kb} T^b_{ij} +T^a_{ib} T^b_{jk}  +T^a_{jb} T^b_{ki} .\label{Rdiff}
\ee
This makes it easy to see that with a torsion-free connection, $R$ and $\t{R}$ are equal, but when the torsion is non-zero,  $R$ and $\t{R}$ can differ.   

The first Bianchi identity is:
\be
\sum_{(i,j,k)} R^a_{kij} =\sum_{(i,j,k)} \(T^a_{bk} T^b_{ij} +  \nabla_i T^a_{jk}\).  \label{bianchi}
\ee
Since in our case the curvature $R$ is zero, (\ref{Rdiff}) and (\ref{bianchi}) give a nice relation between the curvature $\t{R}$ and the torsion: 
\be
\t{R}^a_{kij} = \nabla_k T^a_{ij}  \qquad \qquad (R^a_{kij}=0). \label{rzerotr}
\ee


%

\end{document}